# Three-Dimensional Dynamic Cutting Model

CONSTANTIN ISPAS[1], CLAUDIU BISU[1], ALAIN GERARD[2], DORU BARDAC[3]
[1] Machines and Production Systems Departament
“POLITEHNICA” University of Bucharest
Splaiul Independenţei no. 313 Street,
ROMANIA
[2] Mechanic and Physic Laboratory
University Bordeaux 1
351 Cours de la Libération, 33405 Talence Cedex
FRANCE
[3] Manufacturing Department
“POLITEHNICA” University of Bucharest
Splaiul Independenţei no. 313 Street,
ROMANIA

ispas1002000@yahoo.fr, cfbisu@gmail.com, 125
alain.gerard@u-bordeaux1.fr, doru@bardac.net

***Abstract:*** The determination of a dynamic law of cut is complex and often very difficult to develop. Several formulations were developed, in very complex ways being given that 3 **A**D crosses from there, the number of variables is much higher than out of orthogonal cut. The existence of the plan of displacements and the correlations with the elastic characteristics of the machining system thus make it possible to simplify the dynamic model 3D. A dynamic model on the basis of experimental approach is proposed. Simulation is in concord with the experimental results.

## 1 Introduction

In order to reduce the costs and the times of adjustment of the manufacturing processes, the model approach seems an ideal solution. To represent as well, as possible, the process of cut, these models must integrate the phenomena physical, thermal and dynamic related on the chip formation and the generation of surfaces. Industrial configurations of machining as well as former research tasks [1], [2] show the need for a three-dimensional modeling of the method, which increases the complexity of the resolution of the problem of the dynamics of the cut. The dynamic model is based on semi-analytical thermomechanical model of three-dimensional cutting [4] which answers the fine description of the contact tool/part/chip. The model takes into account various areas of stresses that is plastic deformation, as in the primary area of shearing, or of the stresses of contact, and this during the process of formation of the chip. Modeling is carried out under the conditions of self-excited vibrations [6], [7], [8].

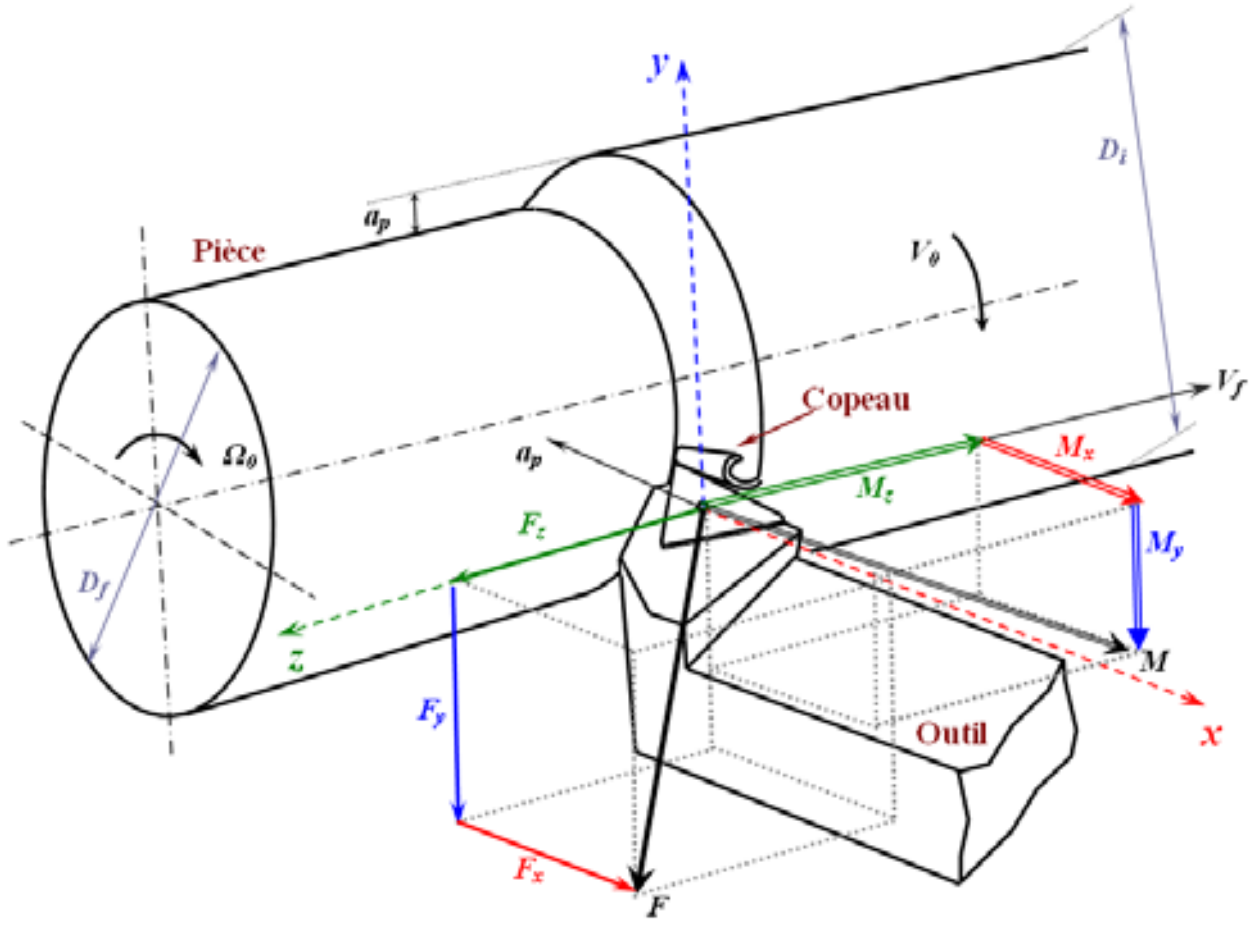

Fig.1.Three dimensional cutting in turning

## 2. Construction of model

The model of cut is based on a configuration 3D of machining, with a system with 3 ddl (degrees of freedom). The movement of the tool is expressed in a reference mark related to the tool which is then projected in the reference mark related to the machining system [10].

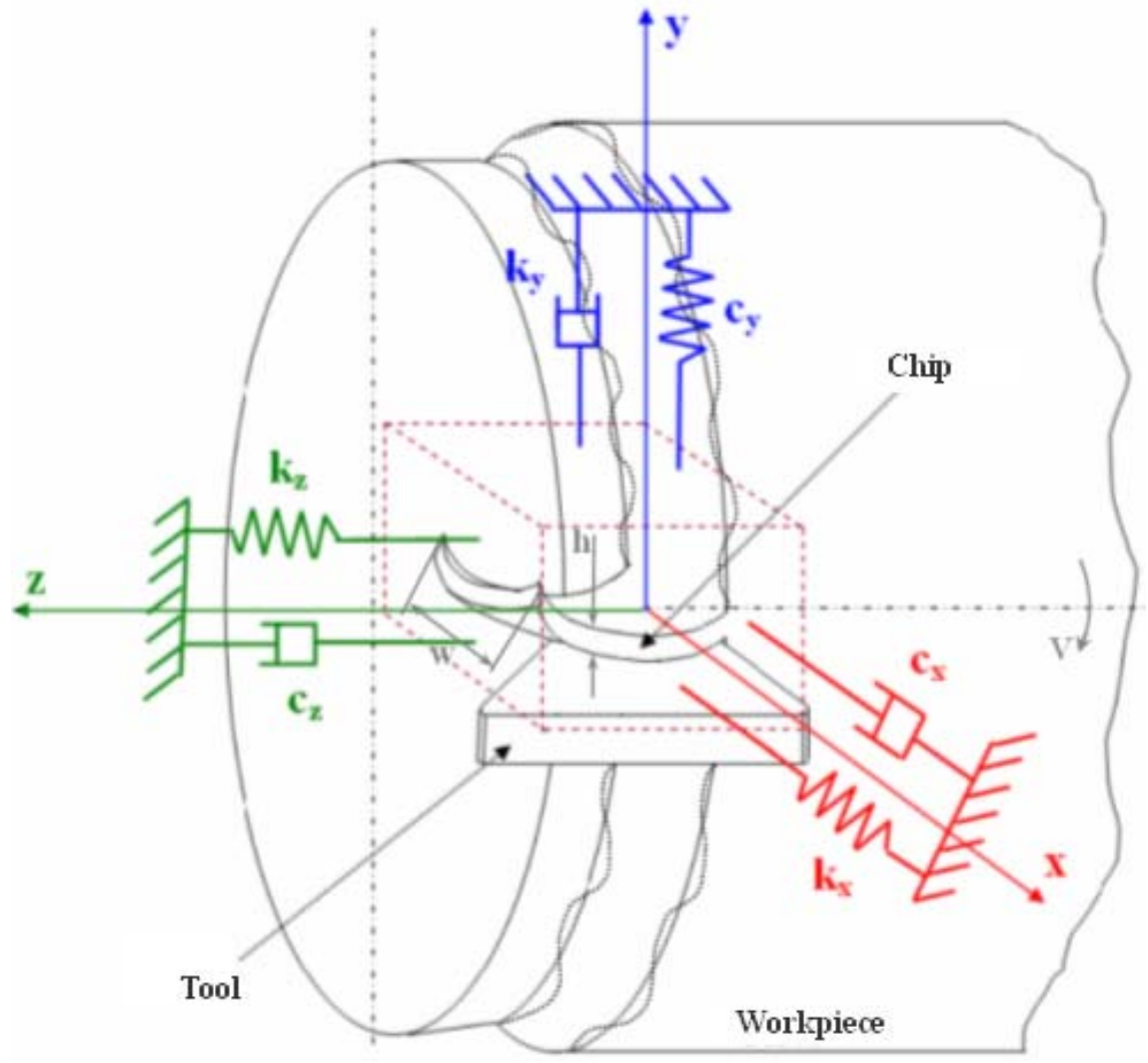

Fig. 2. Dynamic model of cut

## Dynamic description of the contact tool/part/chip

By assumption, the dynamic law of cut defined with rate is established [9]. The transient states are in general to dependant on the initial conditions so that the results of simulation can be compared quantitatively with those of the experiment. With this method, we determine the relation between the cutting forces and the instantaneous parameters of machining. The principle consists in determining the cutting forces starting from the angle optimization of shear integrated in the dynamic model. These contact variations tool/part/chip are due to relative displacements tool/part, tool/chip, and generate variations of the section of chip, speed of the chip, rake angle, clearance angle and shearing angle.

The contact variations tool/part/chip are then examined, on each direction. The relative displacement of the tool causes certain physical variations [3]**.** The tool tip displacements plane *Pu* is characterized by its normal vector noted **nu** (Fig.3). The characterization of this plane by its normal is carried out for each test depending on the feed motion in turning. We note that the normal component according to the feed motion direction almost does not evolve according to this one. On the other hand, a very light decrease (respectively growth) according to the radial direction (respectively, the cutting axis) is observed in dependence with the feed motion values. These variations are, however, so weak that they do not affect the standard of the plane direct normal, which thus can be considered constant according to the feed motion. The existence and the determination of this plane provide simplifying important informations about the configuration to be adopted in order to write a semianalytical model that is under development . In particular, we will further see that this plane allows to bring back the cutting three-dimensional problem, with three directions displacements, using a simpler plane model that is inclined compared to axes (*x*, *y*, *z*) related to the machine tool.

# 3 Theoretical model

Dynamic modeling is based on the semi-analytical model developed by [5], which provide us the face values of the results at the time of an iteration. In this model it is possible to know the forces compared to a length corresponding to each area of stress.

## 3.1 Three dimensional model

The determination of the parameters of contact and the dynamic geometrical parameters constitutes the first stage which leads to the evaluation of the cutting forces.

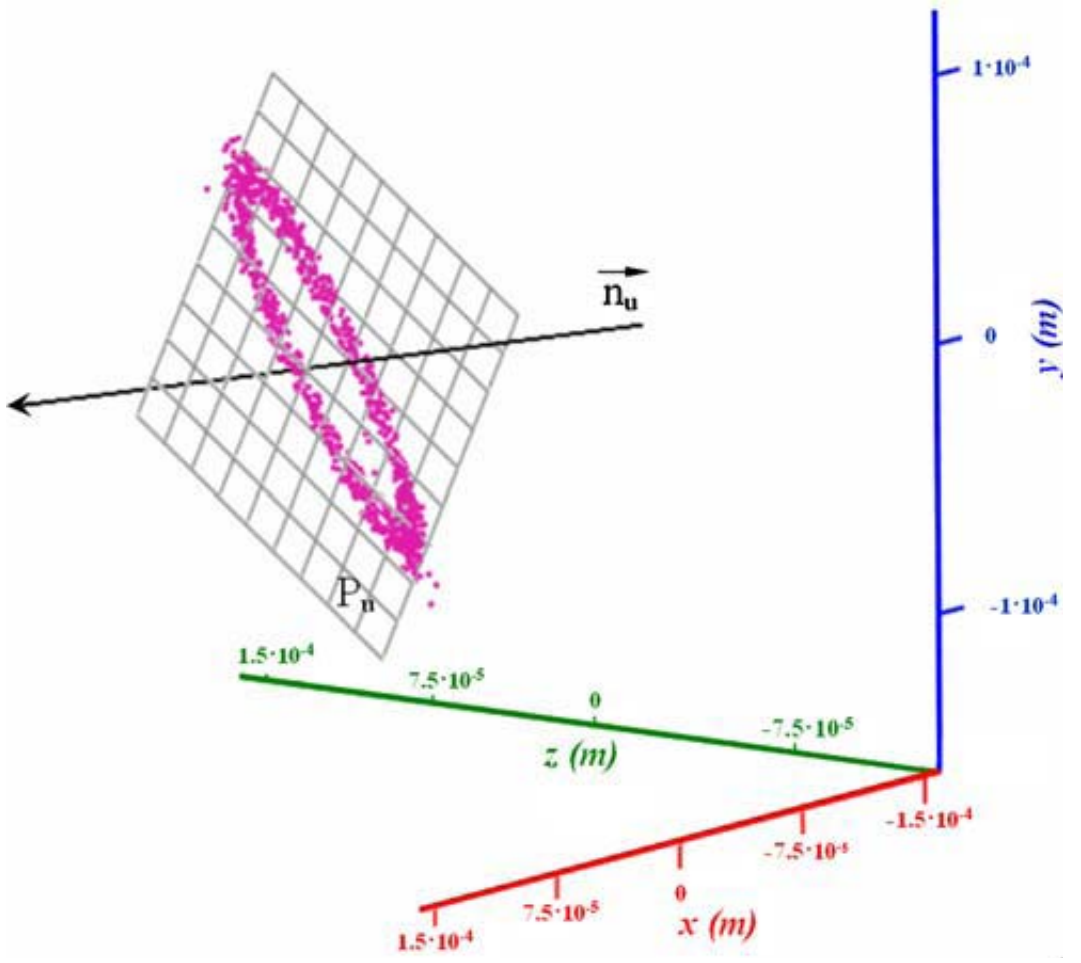

Fig.3. Tool tip displacements plane *Pu*

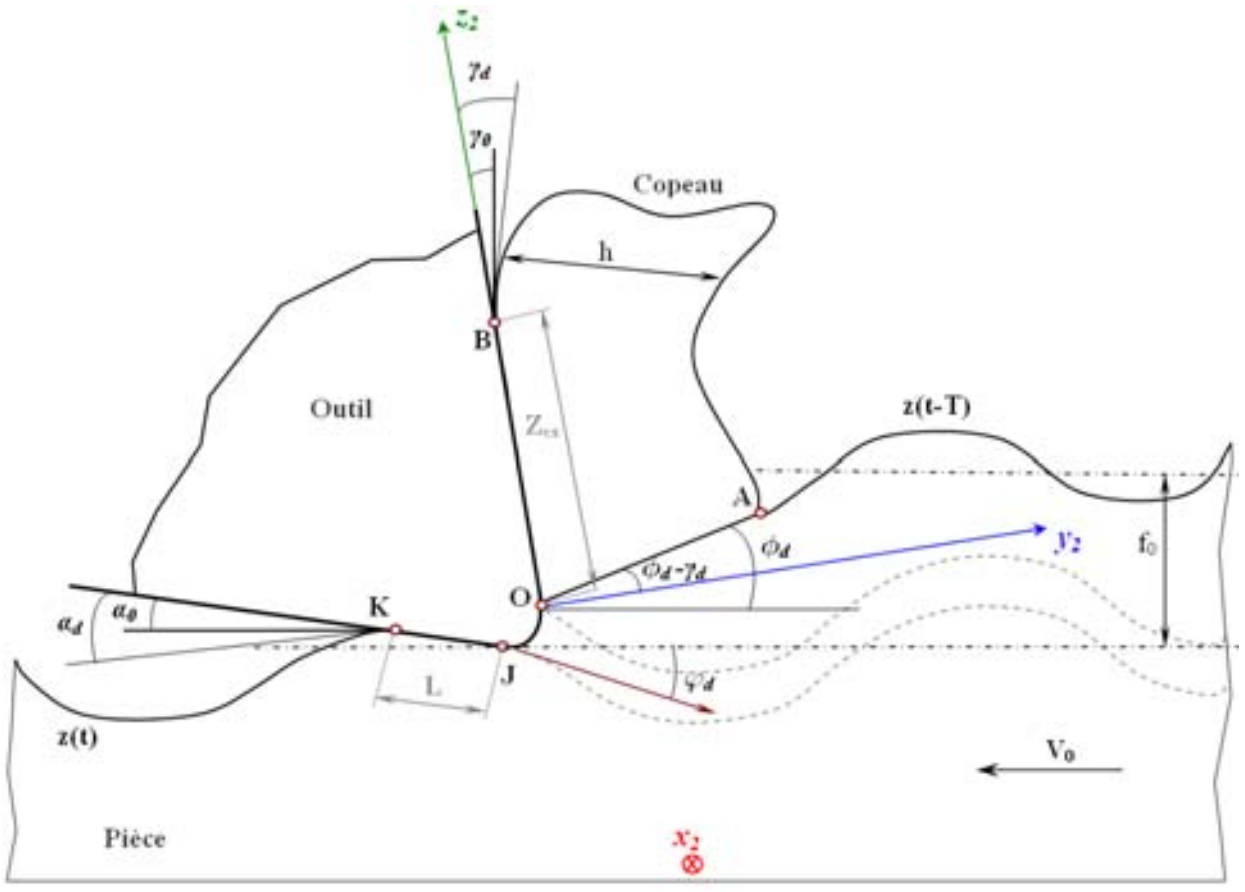

Fig. 4. Description of the variation of contact tool/part/chip.

● in the secondary area of shearing we have (OB)

● the area of skin is divided into two parts: the area of the radius of acuity (area OJ) and the rectilinear area (area *JK*), Projection in the reference mark $x_2$, $y_2$, $z_2$ of the cutting forces (Fig.3) and then passage in the reference mark total makes it possible to write the differential connection in the matrix shape of three-dimensional mode:

$$[M_3]\cdot\ddot{\vec{u}}_3(t)+[C_3]\cdot\dot{\vec{u}}_3(t)+[K_3]\cdot\vec{u}_3(t)=\vec{F}_3(t) \quad (1)$$

with $[M_3]$ the matrix masses, $[C_3]$ the matrix damping and $[K_3]$ the matrix of stiffness. Here $u_3(t)$ and $F_3(t)$ are respectively the vector displacement and the vector forces in the three-dimensional reference coordinates.

### 3.2 Two-dimensional dynamic model after projection

The experimental analysis showed the existence of a specific plan of tool displacement, in that the tool point describes an ellipse [6]. Displacements are generated during the variation of the cutting forces which are located on a level equivalent to the plan of displacements.

The determination of this plan enables us to adopt a real configuration of the cut, in this new reference mark of axes related to the ellipse axes in the plan described. This resulted in passing to a two-dimensional model are equivalent to characterize the three-dimensional cutting. A plan different from the orthogonal plan of cutting machining but characterizing the dynamic behavior of the machining system. At the time of this passage, with the determination of the new reference marks ($n_{fa}$, $n_{fb}$), we modelize the cutting forces in the two-dimensional reference mark (Fig. 4). Thanks to the basic change we can as brought back displacements in the same reference mark as the forces.

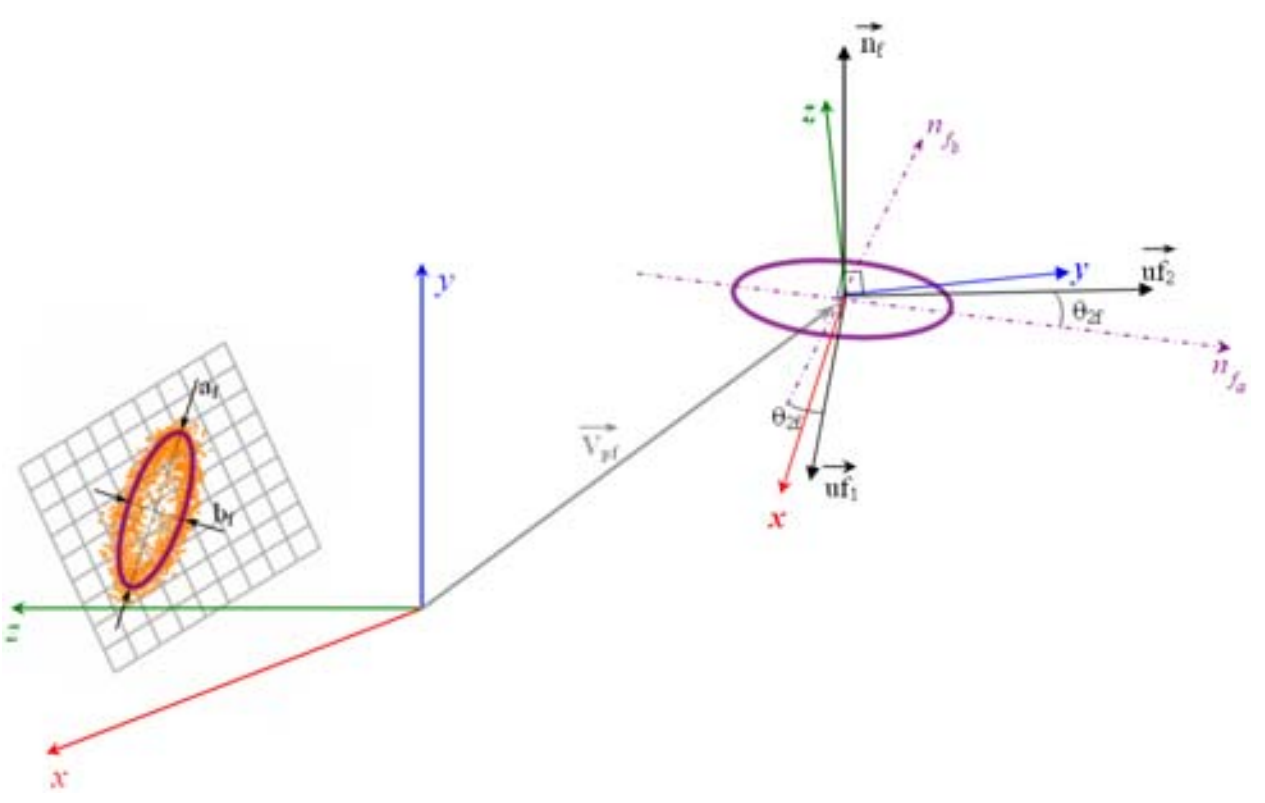

Fig.5. Passage in 2D.

Then the system can be written in the matrix form :

$$[M_2]\cdot\ddot{\vec{u}}_2(t)+[C_2]\cdot\dot{\vec{u}}_2(t)+[K_2]\cdot\vec{u}_2(t)=\vec{F}_2(t) \quad (2)$$

with $[M_2]$ the mass matrix, $[C_2]$ the damping matrix and $[K_2]$ the stiffness matrix; $u_2(t)$ indicates the vector position and $F_2(t)$ the force vector components according to displacement and the force vector components according to the two directions determined during projection 2D.

## 4 Results and analyzes

The signals treatment of the three-direction accelerometers gives, by integration, displacements according to the three directions. An example is presented in the Fig. 6 related to a cutting depth of 2 mm. For these tested conditions, the system is stable; the signal has very low amplitudes, on the order of micrometers; and the chip section remains constant. The part surface quality is correct, with a total roughness *(Rt)* of 4.3 µm. The analysis of measurements for the depth of 3 mm cut give similar results; the displacement amplitudes are in the same region, lower than 10 µm, and the chip section does not vary, this one has the same continuity characteristics as for ap = 2 mm, but very light undulations appear on the workpiece surface. Actually, the characteristics start to change distinctly from the depth of 4 mm cut. The tool point displacement amplitudes increase and the signal sinusoidal character has high amplitude compared to the previous cases. The system starts to become unstable. The chip develops important periodic undulations and significant section variations appear.

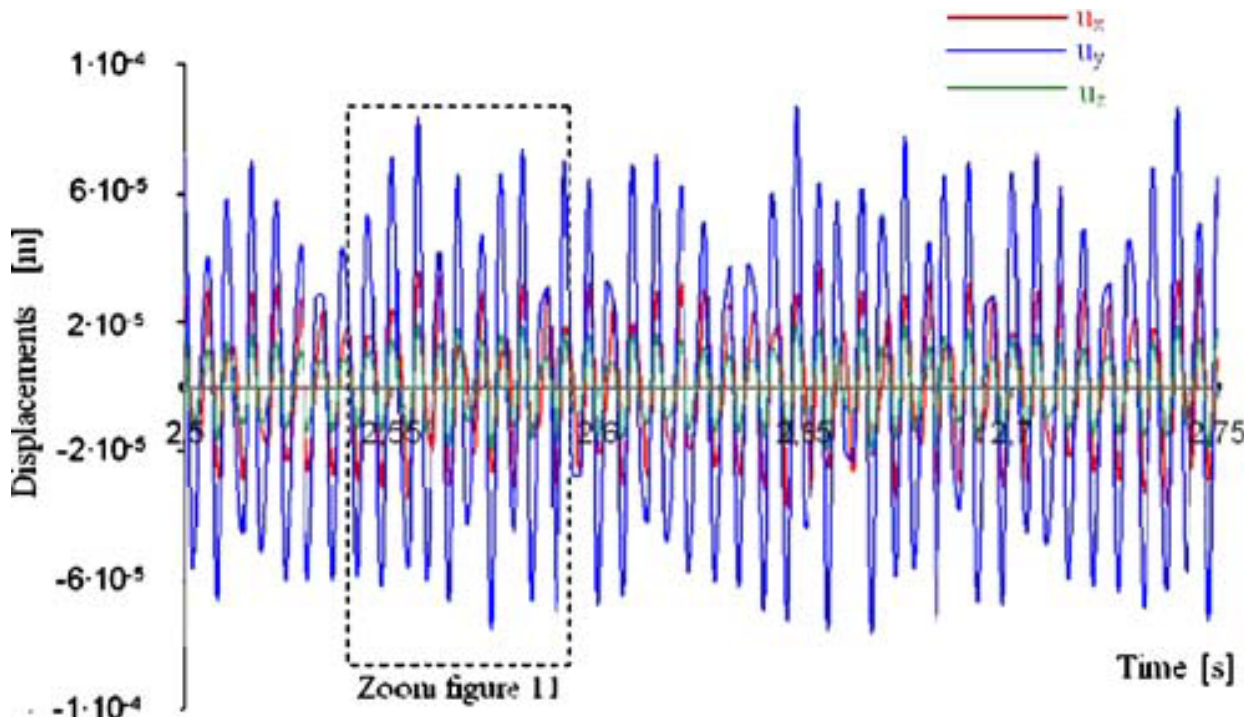

Fig.6. The tool tip displacements on the three *x*, *y*,and *z* directions related to the machine tool and considering ap = 5 mm, *f* = 0.1 mm/rev, *N* = 690 rpm

All chips are type 1.3 (ISO 3685). The machined surface part also has weak undulations. The increase of cutting depth until 5 mm allows to reach a mode of totally unstable cutting and the self-excited vibration appearance is clearly identified. The two systems of equations are solved separately using the method of Runge Kutta order 5 applied to the differential equations resolution. The results obtained by simulation in the two-dimensional case are coherent with the experimental data (Table 1 and figure 7):

*Table 1.*

Comparison experiment/simulation.
where Fa and Fb represent the large axis and the small axis of the ellipse, in the theoretical case and the experimental case.

| Fa (expérimental) | Fa (théorique) | Fb (expérimental) | Fb (théorique) |
|---|---|---|---|
| 690 N | 713 N | 280 N | 228 N |

## 5. Conclusions

A model of dynamic cut original was developed by the laboratories of Bordeaux: *LMP* and *LGM ² B* (University Bordeaux 1). It integrates the stationary model of cut developed While made it possible to transform the writing of the problems 3D into problems 2D, faster and more easily to solve It represents a first evolution of a three-dimensional dynamic cutting model. The model allows to obtain results in concord with the experimental values. the prospects offered by this model are important currently making it possible to predict the dynamic forces of cutting but also the morphology. reviewing some existing dynamic phenomena at the time of the vibratory cut, and by assumptions put forth on this modeling, a dynamic model is designed.

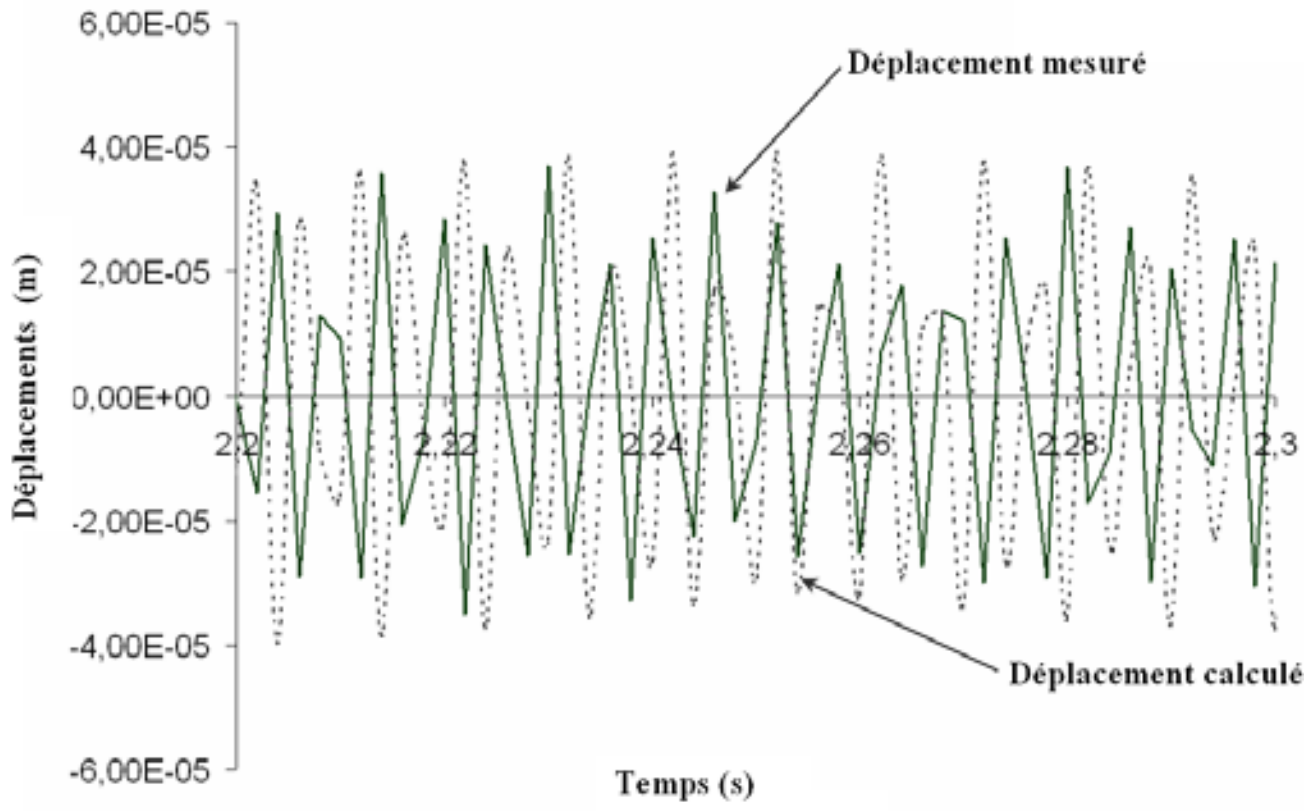

Fig.7: Simulation results of the two-dimensional dynamic cutting model

A first resolution of the system is carried out for which the results obtained are coherent.

The existence and the analysis of the plan of displacements it is also important to note for this research the plane determination attached to the tool tip displacements, which is carried out according to an ellipse, the surface of which is an increasing function depending on the feed motion value; this aspect is in perfect coherence with the power level injected in the system during machining.